\documentclass[preprint,12pt]{article}
 
\usepackage{amssymb}
\usepackage{caption}
\usepackage{breqn}
\usepackage{graphicx}
\usepackage{amsmath} 
\usepackage{cite}

\usepackage{xparse}
\ExplSyntaxOn
\NewDocumentCommand{\mref}{m}{\quinn_mref:n {#1}}
\seq_new:N \l_quinn_mref_seq
\cs_new:Npn \quinn_mref:n #1
 {
  \seq_set_split:Nnn \l_quinn_mref_seq { , } { #1 }
  \seq_pop_right:NN \l_quinn_mref_seq \l_tmpa_tl
  ( 
  \seq_map_inline:Nn \l_quinn_mref_seq
    { \ref{##1},\nobreakspace } 
  \exp_args:NV \ref \l_tmpa_tl 
  ) 
 }
\ExplSyntaxOff

\begin{document}

\begin{center}

\textbf{Real valued functions for BFKL eigenvalue}

\small
Mohammad Joubat$^{(a)}$   and Alex Prygarin$^{(b)}$  
\\
$^{(a)}$ Department of Mathematics, Ariel University, Ariel 40700, Israel\\
$^{(b)}$ Department of Physics, Ariel University, Ariel 40700, Israel 
\end{center}
\normalsize

\normalsize

\begin{abstract}
We consider   known expressions for the  eigenvalue of the Balitsky-Fadin-Kuraev-Lipatov (BFKL) equation in $N=4$ super Yang-Mills theory as a real valued function of two variables. We define new real valued functions of two complex conjugate variables that have a definite complexity analogous to the weight of the nested harmonic sums. We argue that those functions span a general  space of functions for the BFKL eigenvalue at any order of the perturbation theory.

\end{abstract}

\newpage

\section{Introduction}
The analytic solution of the Balitsky-Fadin-Kuraev-Lipatov~(BFKL)~\cite{BFKL} equation is traditionally represented as the  eigenvalue problem in the space of real  Mellin coordinates of   conformal spin $n$ and anomalous dimension $\nu$. Those two variables are real, but the known leading order~(LO) and the next-to-leading order~(NLO) BFKL~\cite{NLO} eigenvalues are   written in terms of   polygamma functions and their generalizations as functions of complex variables $z=-\frac{1}{2}+ i \nu +\frac{|n|}{2}$ and $\bar{z}=-\frac{1}{2}- i \nu +\frac{|n|}{2}$.
 The overall expression is real due to the cancellation of the imaginary part and it would be more natural to use real valued functions to describe the real valued BFKL eigenvalue. As a starting point we use the NLO BFKL eigenvalue in $N=4$ SYM derived from the corresponding QCD expression by Kotikov and Lipatov~\cite{KotikovDGLAP1, KotikovDGLAP2} due to the uniform complexity of the functions that build it.
 The uniform complexity of NLO BFKL eigenvalue in $N=4$ SYM is related to the principle of the maximal transcendentality formulated by Kotikov and Lipatov~\cite{maxtrans}, which can be used as a practical tool in  loop calculations.   
 
 The recent results for higher order corrections to the BFKL eigenvalue stem from integrability  techniques use nested harmonic sums as a convenient space of functions. The nested harmonic sums are functions of one variable and the currently available results for higher order corrections to the BFKL eigenvalue  are given in the analytic form as functions of either $\nu$ or $n$, but not both of them.   The nested harmonic sums represent a convenient choice of possible space of functions mostly due to their pole structure that coincides with that of polygamma functions and their generalizations, as well as, because of the fact that they are properly labeled with so-called \emph{weights} that allow to identify the complexity of expression. The maximal  complexity of functions in the BFKL eigenvalue depends on the loop order and gives a connection between the corresponding expressions in QCD and N=4 SYM due the principle of maximal transcedentality. 
 In this paper we continue to investigate the analytic properties of the BFKL eigenvalue and related special functions~\cite{Bondarenko:2015tba, Bondarenko:2016tws, Prygarin:2018tng, Prygarin:2018cog, Prygarin:2019ruv, Joubat:2019esj, Joubat:2020hvc, Joubat:2020vrw}. We put an emphasis on special functions   that span the space of functions relevant for the BFKL eigenvalue and are universal at any order of the perturbation theory. In this paper we propose   real valued functions of two  complex conjugate variables as a candidate to define the proper space of functions for this purpose. 
The new  real valued function combine the useful features of the nested harmonic sums and generalized polygamma functions such as a well defined complexity~(weight) and the correct pole structure.  

The paper is organized as follows. In the next section we analyze the functions that appear in  LO and  NLO expressions of the  BFKL eigenvalue in $N=4$ SYM. Then we show that the special real valued combinations of those functions possess common functional form of a well defined complexity~(i.e. weight). In the last section we list all possible real valued functions at weight one and  two. In the Appendix we provide a full list of the new real valued functions at weight three.

\section{BFKL eigenvalue}
In this section we consider the  BFKL eigenvalue in $N=4$ Super Yang-Mills~(SYM) theory and the leading order~(LO) and the next-to-leading~(NLO) order that was calculated directly from Feynman diagrams. Then we extend our analysis to the known  higher order corrections to the BFKL eigenvalue obtained using integrability  techniques.   This LO and NLO expressions can be written as follows (see paper by Kotikov and Lipatov~\cite{KotikovDGLAP1,KotikovDGLAP2} )
\begin{eqnarray}\label{omega}
\omega= 4 \bar{a} \left[\chi(n , \gamma)+\bar{a} \; \delta(n, \gamma)\right]
\end{eqnarray} 
where $\gamma= \frac{1}{2}-i \nu$ and 
\begin{eqnarray}\label{chi}
\chi(n, \gamma)= 2 \psi(1) -\psi\left(M\right)-\psi\left(1-\bar{M}\right), 
\end{eqnarray}
\begin{eqnarray}\label{delta}
\delta(n, \gamma)= \phi\left(M\right)+\phi\left(1-\bar{M}\right)-\frac{\omega_0}{2 \bar{a}} \left(\rho\left(M\right)+\rho\left(1-\bar{M}\right)\right)
\end{eqnarray}
in terms of
\begin{eqnarray}
M=\gamma+\frac{|n|}{2}, \;\;\;   \bar{M}= \gamma-\frac{|n|}{2}.
\end{eqnarray} 
Here $\psi(z)$ is the Euler $\psi$-function defined in through  the logarithmic derivative of the Gamma function  $\psi(z)=\frac{d \ln \Gamma(z)}{dz}$. The coupling $\bar{a}=\frac{g^2 N_c}{16 \pi^2}$ is given in terms of the coupling constant $g$ in the DREG scheme.
The functions $\rho(M)$ and $\phi(M)$ are give by 
\begin{eqnarray}
\rho(M) = \beta'(M)+\frac{1}{2} \zeta(2)
\end{eqnarray}   
and 
\begin{eqnarray}
\phi(M)= 3 \zeta(3) +\psi''(M)-2 \Phi_2(M)+2 \beta'(M)\left( \psi(1)-\psi(M)\right)
\end{eqnarray}
where 
\begin{eqnarray}
\beta'(M) =\frac{1}{4} \left[ \psi'\left(\frac{M+1}{2}\right)- \psi'\left(\frac{M}{2}\right) \right]=-\sum_{r=0}^{\infty} \frac{(-1)^r}{(M+r)^2}
\end{eqnarray}
and 
\begin{eqnarray}
\Phi_2 (M)= \sum_{k=0}^{\infty} \frac{\beta'(k+1)}{k+M}+\sum_{k=0}^{\infty} \frac{(-1)^k\psi'(k+1)}{k+M}-\sum_{k=0}^{\infty} \frac{(-1)^k\left( \psi(k+1)-\psi(1)\right)}{(k+M)^2}.
\end{eqnarray}

The expressions of $\chi(n,\gamma)$ and $\delta(n,\gamma)$ are real functions of $\gamma= \frac{1}{2}-i \nu$ and  for  real $\nu$ and $n$. They are built of polygamma functions and their generalizations, which are complex functions of one complex variable. They appear in $\chi(n,\gamma)$ and $\delta(n,\gamma)$ in particular combinations that cancel the imaginary part making the overall expression being  real for real $\nu$ and $n$. In the next section we analyze those special combinations and define a real-valued functions of $\nu$ and $n$.

\section{Definitions}
In this section we analyze the real valued combinations of the complex functions of one variable that build the BFKL eigenvalue. Based on this analysis we define the real-valued functions of two real variables with a definite complexity  that matches   the weight of the corresponding nested harmonic sums. 

The nested harmonic sums are defined~\cite{HS1,Verm1998uu,Blum1998if,Rem1999ew} in terms of  nested summation for $n\in \mathbb{N}$
\begin{eqnarray}\label{defS}
S_{a_1,a_2,...,a_k}(n)=  \sum_{n \geq i_1 \geq i_2 \geq ... \geq i_k \geq 1 }   \frac{\mathtt{sign}(a_1)^{i_1}}{i_1^{|a_1|}}... \frac{\mathtt{sign}(a_k)^{i_k}}{i_k^{|a_k|}}.
\end{eqnarray}
The harmonic sums  are defined for  real integer values of $a_i$~(excluding zero), which  build the alphabet of the possible negative and positive indices uniquely  labeling $S_{a_1,a_2,...,a_k}(n)$.  
In the definition of nested harmonic sums $S_{a_1,a_2,...,a_k}(n)$ in~\mref{defS} $k$ is  the depth and $w=\sum_{i=1}^{k}|a_i|$ is the weight. The nested harmonic sums are defined for positive integer values of the argument and require analytic continuation to the complex plane done using their integral representation.  

In our analysis of the functions in the BFKL eigenvalue we start with the digamma functions defined in terms of the logarithmic derivative of the Gamma function as follows
\begin{eqnarray}
\psi(z)\equiv \frac{d \ln \Gamma(z)}{dz }.
\end{eqnarray}
For the purpose of the present discussion we use the series representation of the digamma function  
\begin{equation}\label{psi0}
 \psi(z)-\psi(1) =\sum_{k=1}^{\infty}\frac{z-1}{k (k+z-1)}.
\end{equation}
where $z$ is a complex variable 
\begin{eqnarray}
z=x+i y
\end{eqnarray}
for real $x$ and $y$. The digamma function appears in the BFKL eigenvalue in \mref{omega} as a linear combination of   arguments $z$ and $\bar{z}$ 
\begin{eqnarray}
\psi(z)+\psi(\bar{z})-2\psi(1)&=& \sum_{k=0}^{\infty} \left(\frac{2}{k+1}-\frac{1}{k+z}-\frac{1}{k+\bar{z}}\right)
\\
&&=\sum_{k=1}^{\infty} \left(\frac{2}{k+1}-\frac{2(k+x)}{(k+x)^2+y^2}\right).
\end{eqnarray}
In a similar way we can reduce the the first derivative of digamma function~(called trigamma function) defined by 
\begin{eqnarray}\label{psi1}
\psi'(z)=\sum _{k=0}^{\infty } \frac{1}{(k+z)^2}.
\end{eqnarray}
The real-valued combination of trigamma function reads
\begin{eqnarray}
\psi'(z)+\psi'(\bar{z})=\sum _{k=0}^{\infty } \left( \frac{1}{(k+z)^2} +\frac{1}{(k+\bar{z})^2}\right)
=2\sum_{k=0}^{\infty} \frac{(k+x)^2-y^2}{ \left((k+x)^2+y^2\right)^2}.
\end{eqnarray}
We proceed with the second derivative of the digamma function and get
\begin{eqnarray}\label{psi2}
\psi''(z)+\psi''(\bar{z})&=&-2\sum _{k=0}^{\infty } \left( \frac{1}{(k+z)^3} +\frac{1}{(k+\bar{z})^3}\right) \nonumber
\\
&&
=-4\sum_{k=0}^{\infty} 
\frac{(k+x)^3-3 y^2 (k+x)}
{ \left((k+x)^2+y^2\right)^3}.
\end{eqnarray}
From the three expressions in \mref{psi0}, \mref{psi1} and \mref{psi2} we can extract a general structure of their  terms  
\begin{eqnarray}\label{prelim}
 \sum_{k=0}^{\infty}\frac{  (k+x)^{\alpha- 2 \beta} y^{2\beta}}
{((k+x)^2+y^2)^{\alpha}},
\end{eqnarray}
where $\alpha$ and $\beta$ are positive integers subject to the following relation $\alpha- 2 \beta \geq 0$. The expression in \eqref{prelim} is a real-valued function of $z$ and $\bar{z}$ with a definite complexity depending on  the  choice of the parameter $\alpha$ that corresponds to the weight of nested harmonic sums. 
 It is important to emphasize that the complexity of the functions in \eqref{prelim} is fully determined by $\alpha$ and  does not depend on the parameter $\beta$, i.e.  is the same for  any positive integer $\beta$   that satisfies the relation  $\alpha- 2 \beta \geq 0$.

Next we analyze more complicated functions $F_i(z)$ that are present in the NLO BFKL eigenvalue
\begin{eqnarray}
&&F_1(z)=\sum_{k=0}^{\infty}\frac{(-1)^k \psi'(k+1)}{k+z},\\
&& F_2(z)=\sum_{k=0}^{\infty} \frac{(-1)^k \left( \psi(k+1)-\psi(1)\right)}{(k+z)^2},
\\
&&
F_3(z)=\sum_{k=0}^{\infty} \frac{\beta'(k+1)}{k+z}, 
\end{eqnarray}
where 
\begin{eqnarray}
\beta'(z)=-\sum_{r=0}^{\infty}\frac{(-1)^r}{(r+z)^2}.
\end{eqnarray}

The functions $F_i(z)$ always appear in the NNLO eigenvalue  \eqref{delta} in the real valued combinations $F(z)+F(\bar{z})$. We start with $F_1(z)$  for $z=x+i y$ and write  
\begin{eqnarray}\label{F1}
F_1(z)+F_1 (\bar{z})&=&\sum_{k=0}^{\infty} (-1)^k \psi'(k+1)\left(\frac{1}{k+z}+\frac{1}{k+\bar{z}}\right) \nonumber
\\
&&=2\sum_{k=0}^{\infty}\frac{(-1)^k(k+x)}{(k+x)^2+y^2}  \psi'(k+1) \nonumber
 \\
&& = 2 \sum_{k=0}^{\infty}\frac{(-1)^k(k+x)}{(k+x)^2+y^2}  \left( -S_2(k)+\frac{\pi^2}{6}\right),
\end{eqnarray}
where we used the analytic continuation of the harmonic sum 
\begin{equation}
\psi'(k+1)=-S_{2}(k)+\frac{\pi^2}{6}.
\end{equation}

Next we consider $F_2(z)$ and its real valued combination
\begin{eqnarray}\label{F2}
&&F_2(z)+F_2(\bar{z})=\sum_{k=0}^{\infty} (-1)^k \left( \psi(k+1)-\psi(1)\right)\left(\frac{1}{(k+z)^2}+\frac{1}{(k+\bar{z})^2}\right)  \nonumber \\
&&
=2\sum_{k=0}^{\infty} (-1)^k \frac{(k+x)^2-y^2}{\left((k+x)^2+y^2\right)^2}S_1(k),
\end{eqnarray}
where   the analytic  continuation for the harmonic sum is given by
\begin{eqnarray}
S_1(k)=\psi(k+1)-\psi(1).
\end{eqnarray}

Finally we repeat the analysis for $F_3(z)$. 

\begin{eqnarray}\label{F3}
F_3(z)+F_3(\bar{z})&=& \sum_{k=0}^{\infty}
\left(\frac{1}{k+z}+\frac{1}{k+\bar{z}}\right)\beta'(k+1) \nonumber
\\
&&
=\sum_{k=0}^{\infty}
\frac{2 (k+x)}{(k+x)^2+y^2}
\left( -\overline{S}^{+}_{-2}(k)+\frac{\pi^2}{12}\right)
\end{eqnarray}
where we used 
\begin{eqnarray}
\beta'(z)=-\sum_{r=0}^{\infty}\frac{(-1)^r}{(r+z)^2}
\end{eqnarray}
and the relation  
\begin{eqnarray}
\beta'(k+1)= -\overline{S}^{+}_{-2}(k)+\frac{\pi^2}{12}
\end{eqnarray}
for the analytic the analytic continuation for the harmonic sum   from even $k$ to natural numbers
\begin{equation}
\overline{S}_{-2}^+(k)= (-1)^k S_{-2}(k)+\left(1+(-1)^k\right)\frac{\pi^2}{12}.
\end{equation}

The $\beta'(z)$ also appears in the NNLO BFKL eigenvalue  in the real valued combination 

\begin{eqnarray}\label{betapcomb}
\beta'(z)+\beta'(\bar{z})= -2\sum_{k=0}^{\infty}(-1)^k \frac{(k+x)^2-y^2}{((k+x)^2+y^2)^2} 
\end{eqnarray}
where $x=\operatorname{Re}(z)$ and $y=\operatorname{Im}(z)$.

From the real valued expression in \mref{psi0},\mref{psi1},\mref{psi2},\mref{F1},\mref{F2},\mref{F3} and \mref{betapcomb} one can read out the general form of the  real-valued functions 
\begin{eqnarray}
\sum_{k=0}^{\infty}\frac{(-1)^{\eta k}(k+x)^\gamma y^{2\beta}}
{((k+x)^2+y^2)^\alpha}  \overline{S}^+_{\{a_i\}}(k)
\end{eqnarray}
where  $\alpha$ and $\beta$ are natural numbers and $\eta=0,1$.   
Here  $\overline{S}^+_{\{a_i\}}(k)$ denotes nested harmonic sum analytically continued from even values of $k$ to natural numbers. The details of this analytic continuation can be found in the paper by Kotikov and Velizhanin~\cite{Velizhanin}. 
 In order to have a function with  a well defined  weight we set  $\gamma+2 \beta=|\alpha$. The weight of the function in this case is given by 
$w=\alpha+\sum_{i=1}^{d} |a_i|$, where $d$ is the depth of the harmonic sum $S_{\{a_i\}}(k)$. For sake of clarity of presentation we can eliminate $\eta$ and set $\alpha $ to be integer number not including zero. 
Thus for a given weight we define the most general form of functions that can appear in the BFKL eigenvalue
\begin{eqnarray}\label{ddef}
D^{\beta}_{\alpha,  \{a_i\}}(z,\bar{z}) \equiv \sum_{k=0}^{\infty}\frac{\mathtt{sign}(\alpha)^{ k} (k+x)^{|\alpha|- 2 \beta} y^{2\beta}}
{((k+x)^2+y^2)^{|\alpha|}} \overline{S}^+_{\{a_i\}}(k),
\end{eqnarray}
where $\alpha$ is an integer number~(not including zero) and $\beta$ is a natural number~(including zero) subject to the condition $|\alpha|- 2 \beta \geq 0$. Then the weight of the  $D^{\beta}_{\alpha,  \{a_i\}}(z,\bar{z})$ function in \mref{ddef} reads
\begin{eqnarray}\label{weight}
w=|\alpha|+\sum_{i=1}^{d} |a_i|.
\end{eqnarray}
The weight in \mref{weight} is equivalent to the weight of the nested harmonic sums of the same complexity.

  The variables   $x=\operatorname{Re}(z)$ and $y=\operatorname{Im}(z)$  are 
$x=\frac{1}{2}+\frac{n}{2}$ and $y=\nu$ in the traditional notation of the BFKL approach.   Because of the similarity to the dispersive integrals we choose to call the new functions in \mref{ddef} the \emph{dispersive functions}. 
 The  dispersive  functions defined in \mref{ddef}   are real-valued functions of two complex conjugate variables $z$ and $\bar{z}$ and properly capture the analytic structure of the BFKL eigenvalue, which has isolated poles at $\nu=\pm i m/2$ for $m \in \mathbb{Z}$.
One of the main features of the  dispersive functions in \mref{ddef} is that they  are defined at definite complexity, which is useful for building functional basis and then fitting the free coefficients based on the known results. This approach is parallel to the one originally used by Gromov, Levkovich-Maslyuk and Sizov~\cite{gromov} in calculating the functional form of the NNLO BFKL eigenvalue in $N=4$ SYM for $n=0$ in terms of the nested harmonic sums~(see also a parallel calculation by Velizhanin~\cite{velizh}). The expressions   for other values of the conformal spin in  other related calculations can be found in \cite{GromovNonzero,huot} and \cite{Caron-Huot:2015bja, Caron-Huot:2020grv, Gardi:2019pmk, Vernazza:2018gyb, Caron-Huot:2017fxr, Alfimov:2020obh}. 
The real valued dispersive functions can be used in other approach of the BFKL physics such as probabilistic approach \cite{kozlov} or Reggeon effective action approach \cite{bond}. 
The nested harmonic sums defined in 
\mref{defS} were chosen for  constructing  the functional basis  primarily because of the convenience of their labeling, which also defines their complexity in terms of the so-called weight. The weight $w$ of nested harmonic sum $S_{a_1,a_2,...}(n)$ is the sum of the absolute values of their indices, namely $w=\sum_i |a_i|$.  In the case of the dispersive functions $D^{\beta}_{\alpha,  \{a_i\}}(z,\bar{z})$ defined in \mref{ddef} the complexity is given by  $|\alpha|+\sum_i |a_i|$ and it is equivalent to the weight $w$ of the nested harmonic sums. The important difference   between  the dispersive functions $D^{\beta}_{\alpha,  \{a_i\}}(z,\bar{z})$ and  the nested harmonic sums is that the latter are complex  functions of one complex variable (after the analytic continuation), while the dispersive functions $D^{\beta}_{\alpha,  \{a_i\}}(z,\bar{z})$ are real valued functions of a complex variable $z$ and its complex conjugate $\bar{z}$.  
There is limited number of $D^{\beta}_{\alpha,  \{a_i\}}(z,\bar{z}) $ functions  at a given  weight as listed below at weight one and two. The list of all possible $D^{\beta}_{\alpha,  \{a_i\}}(z,\bar{z}) $ can be found in the Appendix.

At weight $w=1$  the functions defined in \mref{ddef} read
\begin{eqnarray}
&& D^0_1(z,\bar{z})=\sum_{k=0}^\infty \frac{(x+k)}{(x+k)^2+y^2}
 \\
 && D^0_{-1}(z,\bar{z})=\sum_{k=0}^\infty \frac{ (-1)^k(x+k)}{(x+k)^2+y^2}
\end{eqnarray}
and at the weight two $w=2$ they are given by
\begin{eqnarray}\label{Dw2}
&& D^0_2(z,\bar{z})=\sum_{k=0}^\infty \frac{(x+k)^2}{\left((x+k)^2+y^2\right)^2}
  \\
&& D^0_{-2}(z,\bar{z})=\sum_{k=0}^\infty \frac{ (-1)^k(x+k)^2}{\left((x+k)^2+y^2\right)^2}
   \\
&& D^0_{1,1}(z,\bar{z}) =\sum_{k=0}^\infty \frac{(x+k)}{ (x+k)^2+y^2 }S_1(k)
   \\
&& D^0_{1,-1}(z,\bar{z})=\sum_{k=0}^\infty \frac{(x+k)}{ (x+k)^2+y^2 }S_{-1}(k)
     \\
&& D^0_{-1,1}(z,\bar{z}) =\sum_{k=0}^\infty \frac{(-1)^k (x+k)}{ (x+k)^2+y^2 }S_1(k)
      \\
&& D^0_{-1,-1}(z,\bar{z})=\sum_{k=0}^\infty \frac{(-1)^k(x+k)}{ (x+k)^2+y^2 }S_{-1}(k)
       \\
&&D^1_2(z,\bar{z})=\sum_{k=0}^\infty \frac{y^2}{\left((x+k)^2+y^2\right)^2}
        \\
&&D^1_{-2}(z,\bar{z}) =\sum_{k=0}^\infty \frac{(-1)^k y^2}{\left((x+k)^2+y^2\right)^2}
\end{eqnarray}

At weight $w=3$ there are $24$ irreducible real valued dispersive functions $D^{\beta}_{\alpha,  \{a_i\}}(z,\bar{z})$ 
\begin{eqnarray}
  &&  \left\{D^0_{1,2}(z,\bar{z}),D^0_{1,-2}(z,\bar{z}),D^0_{-1,2}(z,\bar{z}), D^0_{-1,-2}(z,\bar{z}),D^0_{1,1,1}(z,\bar{z}), D^0_{1,-1,1}(z,\bar{z}), \nonumber
\right.  \\
  &&  D^0_{1,1,-1}(z,\bar{z}), D^0_{1,-1,-1}(z,\bar{z}),D^0_{-1,1,1}(z,\bar{z}), D^0_{-1,-1,1}(z,\bar{z}), \nonumber
   D^0_{-1,1,-1}(z,\bar{z}),\\
  &&D^0_{-1,-1,-1}(z,\bar{z}),   D^0_{3}(z,\bar{z}),D^0_{-3}(z,\bar{z}), D^0_{2,1}(z,\bar{z}),D^0_{2,-1}(z,\bar{z}),D^0_{-2,1}(z,\bar{z}),\nonumber 
  \\
  && D^0_{-2,-1}(z,\bar{z}), 
   D^1_{3}(z,\bar{z}),D^1_{-3}(z,\bar{z}),D^1_{2,1}(z,\bar{z}),D^1_{2,-1}(z,\bar{z}),D^1_{-2,1}(z,\bar{z}), \nonumber
    \\ 
  && \left.D^1_{-2,-1}(z,\bar{z}) \right\}
\end{eqnarray}
and they are given in the Appendix.

The number  of  irreducible $D^{\beta}_{\alpha,  \{a_i\}}(z,\bar{z})$ functions at a given weight  should be compared to the number nested harmonic sums of one variable at the same weight. At weight $w=1$ there are only two $D^{\beta}_{\alpha,  \{a_i\}}(z,\bar{z})$ functions as well as the harmonic sums $S_{1}(z)$ and  $S_{-1}(z)$. At weight $w=2$ there are eight $D^{\beta}_{\alpha,  \{a_i\}}(z,\bar{z})$ functions compared to six independent nested harmonic sums of the linear basis
\begin{eqnarray}
\left\{S_{2}(z), S_{-2}(z), S_{1,1}(z), S_{1,-1}(z),S_{-1,1}(z),S_{-1,-1}(z)
\right\}.
\end{eqnarray}
At weight $w=3$ there are $24$ linearly independent irreducible $D^{\beta}_{\alpha,  \{a_i\}}(z,\bar{z})$ functions compared to $18$ independent nested harmonic sums of the linear basis
\begin{eqnarray}
&&\left\{S_{3}(z), S_{-3}(z), S_{1,2}(z),S_{1,-2}(z),S_{-1,2}(z),S_{-1,-2}(z), S_{2,1}(z),S_{2,-1}(z),\right. \nonumber
\\
&&\left. S_{-2,1}(z), S_{-2,-1}(z),S_{1,1,1}(z),S_{1,1,-1}(z),S_{1,-1,1}(z),S_{1,-1,-1}(z),S_{-1,1,1}(z),\right. \nonumber
\\
&&\left. S_{-1,1,-1}(z),S_{-1,-1,1}(z),S_{-1,-1,-1}(z)
\right\}.
\end{eqnarray}

It is worth emphasizing that for pure real $z$, i.e. for $y=0$ the number of the irreducible $D^{\beta}_{\alpha,  \{a_i\}}(z,\bar{z})$ functions and the nested harmonic sums coincide at weight $w=2$ and $w=3$.  The number of $D^{\beta}_{\alpha,  \{a_i\}}(z,\bar{z})$ functions is larger than the number of the linearly independent nested harmonic sums at a given weight that were used to define   $D^{\beta}_{\alpha,  \{a_i\}}(z,\bar{z})$ functions through real valued pairwise combinations $S_{\{a_i\}}(z-1)+S_{\{a_i\}}(\bar{z}-1)$. This fact implies that the space of $D^{\beta}_{\alpha,  \{a_i\}}(z,\bar{z})$ function is larger than the space of harmonic sums and includes functions that cannot be  expressed in terms of real valued combinations  $S_{\{a_i\}}(z-1)+S_{\{a_i\}}(\bar{z}-1)$. Same holds for the  non-linear basis of nested harmonic sums, where one makes use of the quasi-shuffle identities for building a full set functions for the  irreducible basis.

\section{Summary and Discussions}

In this paper we define new real valued functions $D^{\beta}_{\alpha,  \{a_i\}}(z,\bar{z})$ of a complex variable $z$ and its complex conjugate $\bar{z}$. The definition stems from the real valued combinations of generalized polygamma functions in the known expressions of the BFKL eigenvalue in $N=4$ super Yang-Mills theory. The functions  $D^{\beta}_{\alpha,  \{a_i\}}(z,\bar{z})$ have analytic structure similar to that of the polygamma functions and its generalizations as well as to the nested harmonic sums analytically continued to the complex plane.  We named   $D^{\beta}_{\alpha,  \{a_i\}}(z,\bar{z})$ functions  the \emph{dispersive functions} because of similarity in their definitions to the  dispersive integrals. The important feature of  $D^{\beta}_{\alpha,  \{a_i\}}(z,\bar{z})$ functions is that they are of definite complexity, which corresponds to  the weight of nested harmonic sums. The complexity or transcendentality of the functions is important for building a functional basis for perturbative calculations, where the complexity is determined by loop order. In the case of the multi-Regge kinematics of the BFKL approach each perturbative order increases the maximal  transcendentality of the final expression by two units. For example, the expression of the BFKL eigenvalue at  the leading order has maximal  transcendentality  one, at the next-to-leading order its maximal  transcendentality  is  three, for the next-to-next-to-leading order the maximal  transcendentality is  five etc. The definite    transcendentality  of $D^{\beta}_{\alpha,  \{a_i\}}(z,\bar{z})$ functions is very convenient for building the finite functional basis, which counts  about four hundred terms at weight five, needed for the next-to-next-to-leading order of the BFKL eigenvalue. 

The number of  $D^{\beta}_{\alpha,  \{a_i\}}(z,\bar{z})$ functions is larger than 
a number of possible real valued combinations of generalized polygamma functions building the LO and NLO BFKL eigenvalue.   This means that $D^{\beta}_{\alpha,  \{a_i\}}(z,\bar{z})$ functions cannot be expressed in terms real valued combinations of generalized polygamma functions of one variable in the form of $S_{  \{a_i\}}(z-1)+S_{  \{a_i\}}(\bar{z}-1)$ . In this paper we provide a complete  list of $D^{\beta}_{\alpha,  \{a_i\}}(z,\bar{z})$ up to weight three.    

 \section{Acknowledgement}\label{}

We are indebted to S.~Bondarenko   for inspiring discussions on the topic. This work is supported in part by the Young Researcher Start-Up Grant.

\section{Appendix}\label{appA}

Here we make an explicit list of the dispersive functions defined in \mref{ddef}
 at weight $w=3$. Their general expression is given by  
\begin{eqnarray}\label{ddefA}
D^{\beta}_{\alpha,  \{a_i\}}(z,\bar{z})=\sum_{k=0}^{\infty}\frac{\mathtt{sign}(\alpha)^{ k} (k+x)^{|\alpha|- 2 \beta} y^{2\beta}}
{((k+x)^2+y^2)^{|\alpha|}} \overline{S}^+_{\{a_i\}}(k).
\end{eqnarray}
At weight $w=3$ they read

\begin{eqnarray}
&& D^0_{1,2}(z,\bar{z})= \sum_{k=0}^\infty \frac{  (k+x)}{ (x+k)^2+y^2 }S_{2}(k)
\\
&& D^0_{1,-2}(z,\bar{z})= \sum_{k=0}^\infty \frac{  (k+x)}{ (x+k)^2+y^2 }S_{-2}(k)
\\
&& D^0_{-1,2}(z,\bar{z})= \sum_{k=0}^\infty \frac{(-1)^k (k+x)}{ (x+k)^2+y^2 }S_{2}(k)
\\
&& D^0_{-1,-2}(z,\bar{z})= \sum_{k=0}^\infty \frac{(-1)^k (k+x)}{ (x+k)^2+y^2 }S_{-2}(k)
\\
&& D^0_{1,1,1}(z,\bar{z})= \sum_{k=0}^\infty \frac{  (k+x)}{ (x+k)^2+y^2 }S_{1,1}(k)
\\
&& D^0_{1,-1,1}(z,\bar{z})= \sum_{k=0}^\infty \frac{(-1)^k (k+x)}{ (x+k)^2+y^2 }S_{-1,1}(k)
\end{eqnarray}
\begin{eqnarray}
&& D^0_{1,1,-1}(z,\bar{z})= \sum_{k=0}^\infty \frac{  (k+x)}{ (x+k)^2+y^2 }S_{1,-1}(k)
\\
&& D^0_{1,-1,-1}(z,\bar{z})= \sum_{k=0}^\infty \frac{  (k+x)}{ (x+k)^2+y^2 }S_{-1,-1}(k)
\\
&& D^0_{-1,1,1}(z,\bar{z})= \sum_{k=0}^\infty \frac{(-1)^k (k+x)}{ (x+k)^2+y^2 }S_{1,1}(k)
\\
&& D^0_{-1,-1,1}(z,\bar{z})= \sum_{k=0}^\infty \frac{(-1)^k (k+x)}{ (x+k)^2+y^2 }S_{-1,1}(k)
\\
&& D^0_{-1,1,-1}(z,\bar{z})=\sum_{k=0}^\infty \frac{(-1)^k (k+x)}{ (x+k)^2+y^2 }S_{1,-1}(k) 
\\
&& D^0_{-1,-1,-1}(z,\bar{z})= \sum_{k=0}^\infty \frac{(-1)^k (k+x)}{ (x+k)^2+y^2 }S_{-1,-1}(k)
\end{eqnarray}

\begin{eqnarray}
&& D^0_{3}(z,\bar{z})= \sum_{k=0}^\infty \frac{  (k+x)^3}{\left((x+k)^2+y^2\right)^3}  
\\
&& D^0_{-3}(z,\bar{z})= \sum_{k=0}^\infty \frac{(-1)^k (k+x)^3}{\left((x+k)^2+y^2\right)^3}
\\
&& D^0_{2,1}(z,\bar{z})= \sum_{k=0}^\infty \frac{  (k+x)^2}{\left((x+k)^2+y^2\right)^2}S_{1}(k)
\\
&& D^0_{2,-1}(z,\bar{z})= \sum_{k=0}^\infty \frac{  (k+x)^2}{\left((x+k)^2+y^2\right)^2}S_{-1}(k)
\\
&& D^0_{-2,1}(z,\bar{z})= \sum_{k=0}^\infty \frac{(-1)^k (k+x)^2}{\left((x+k)^2+y^2\right)^2}S_{1}(k)
\\
&& D^0_{-2,-1}(z,\bar{z})= \sum_{k=0}^\infty \frac{(-1)^k (k+x)^2}{\left((x+k)^2+y^2\right)^2}S_{-1}(k)
\end{eqnarray}

\begin{eqnarray}
&& D^1_{3}(z,\bar{z})= \sum_{k=0}^\infty \frac{  (k+x)y^2}{\left((x+k)^2+y^2\right)^3}  
\\
&& D^1_{-3}(z,\bar{z})= \sum_{k=0}^\infty \frac{(-1)^k (k+x) y^2}{\left((x+k)^2+y^2\right)^3}
\\
&& D^1_{2,1}(z,\bar{z})= \sum_{k=0}^\infty \frac{  y^2}{\left((x+k)^2+y^2\right)^2}S_{1}(k)
\\
&& D^1_{2,-1}(z,\bar{z})= \sum_{k=0}^\infty \frac{  y^2}{\left((x+k)^2+y^2\right)^2}S_{-1}(k)
\\
&& D^1_{-2,1}(z,\bar{z})= \sum_{k=0}^\infty \frac{(-1)^k y^2}{\left((x+k)^2+y^2\right)^2}S_{1}(k)
\\
&& D^1_{-2,-1}(z,\bar{z})= \sum_{k=0}^\infty \frac{(-1)^k y^2}{\left((x+k)^2+y^2\right)^2}S_{-1}(k)
\end{eqnarray}

There are $24$   functions $D^{\beta}_{\alpha,  \{a_i\}}(z,\bar{z})$ at weight $w=3$.  Here we use a compact notation for  the nested harmonic sums $S_{\{a_i\}}(k)$ that denote the corresponding sums  $ \overline{S}^+_{\{a_i\}}(k)$ that are analytically continued from even  positive values of the argument to natural numbers using the prescription of Kotikov and Velizhanin~\cite{Velizhanin}.


\end{document}